\pdfoutput=1
\documentclass{article}

\usepackage{arxiv}
\usepackage[numbers,sort&compress]{natbib}
\usepackage[utf8]{inputenc}
\usepackage[T1]{fontenc}
\usepackage{amsmath,amssymb,amsfonts}
\usepackage{bm}
\usepackage{booktabs}
\usepackage{graphicx}
\usepackage{siunitx}
\usepackage{xcolor}
\usepackage{microtype}
\usepackage{xurl}
\usepackage[hidelinks]{hyperref}

\graphicspath{{figures/}}

\newcommand{\C}{\mathbb{C}}

\newcommand{\rvalue}{r}
\newcommand{\Nleaf}{N}                       
\newcommand{\Fb}{\overline{\bm F}}           
\newcommand{\Pb}{\overline{\bm P}}           
\newcommand{\Lb}{\overline{\bm{\mathsf L}}} 
\newcommand{\Td}{\hat{T}^{d}}                

\newcommand{\tridotT}{
  \overset{
    \smash{
      \lower0.6ex\hbox{$
        \mkern2mu
        \substack{
          \textstyle\cdot\phantom{\,\cdot}\\[-0.6ex]
          \textstyle\cdot\,\textstyle\cdot
        }
        \mkern16mu
      $}
    }
  }{T_{l}^{\mu\nu}}
}
\newcommand{\tridotTr}{
  \overset{
    \smash{
      \lower0.3ex\hbox{$
        \substack{
          \textstyle\cdot\phantom{\,\cdot}\\[-0.6ex]
          \textstyle\cdot\,\textstyle\cdot
        }
      $}
    }
  }{T_{r}}
}

\title{FE--ODMN: A two-scale crystal-plasticity framework for AA6022-T4 aluminum sheet deep drawing}

\author{%
  Ting-Ju Wei \\
  Special Melted Products Ltd \\
  President Way, Sheffield S4 7UR, United Kingdom \\
  \And
  Guo-Chi Li \\
  Department of Civil Engineering \\
  National Taiwan University \\
  Taipei 10617, Taiwan \\
  \And
  Yi-Ting Lu \\
  Department of Civil Engineering \\
  National Taiwan University \\
  Taipei 10617, Taiwan \\
  \AND
  Chuin-Shan Chen\thanks{Corresponding author.} \\
  Department of Civil Engineering \\
  Department of Materials Science and Engineering \\
  National Taiwan University \\
  No.~1, Sec.~4, Roosevelt Road, Taipei 10617, Taiwan \\
  \texttt{dchen@ntu.edu.tw} \\
}

\begin{document}
\maketitle

\begin{abstract}
Crystallographic texture governs plastic anisotropy and earing in the deep drawing of aluminum sheet, yet full-field crystal plasticity remains too costly for component-scale simulations. This study presents a two-scale framework, FE--ODMN, in which an orientation-aware deep material network (ODMN) acts as a reduced-order representative volume element (RVE) at every integration point of a finite-element (FE) simulation. The macroscopic response is obtained directly from this RVE, and no macroscopic constitutive model is needed. The network represents the polycrystal with 16, 64, or 256 crystal-plasticity material nodes whose orientations and weights are obtained by generalized spherical harmonics texture compaction. Its interaction directions are trained on linear-elastic homogenization data only, and the crystal-plasticity law at the nodes gives the nonlinear, history-dependent response without retraining. The framework is applied to the cylindrical cup drawing of AA6022-T4 sheet. At the RVE level, the networks reproduce full-field spectral solutions under uniaxial tension along seven in-plane directions and tension--compression reversals, at three to four orders of magnitude lower cost. The error in the Lankford coefficients decreases with network depth, mainly from 64 to 256 nodes. In the cup-drawing simulations, with contact parameters calibrated using the 64-node network, the 256-node network reproduces the measured peak punch force, minimum wall thickness, and cup-height range within 1\,\%, 4\,\%, and 1\,\%. The four ears form at the measured positions, but those along the rolling direction are overpredicted. The local textures change strongly near the cup rim and little at the cup center. A complete simulation takes 1\,400 to 26\,000 core-hours.
\end{abstract}


\section{Introduction}
\label{sec:intro}

Deep drawing forms flat sheet-metal blanks into three-dimensional components through large plastic deformation. In rolled aluminum sheets, the crystallographic texture makes plastic flow direction-dependent, which controls how the material is drawn in and where ears form around the cup rim. The earing profile, the wall-thickness distribution, and the punch force together test any material model for sheet-metal forming. The cup-drawing experiments on AA6022-T4 sheet reported by Tian et al.~\cite{tian2017earing} are a well-documented benchmark that links the measured anisotropy of the sheet to its forming behavior at the component scale. Earing is most sensitive to planar plastic anisotropy, whereas wall thinning and punch force also depend on strain hardening, contact friction, and tooling geometry. Predicting all three therefore requires an accurate description of anisotropic plasticity.

Phenomenological anisotropic plasticity models sheet-metal deformation efficiently through yield criteria and flow rules calibrated against macroscopic measurements~\cite{hill1948theory,barlat2003plane,barlat2005linear}. Its accuracy depends on the available calibration data and on how the evolution of anisotropy is described. Crystal plasticity instead describes plastic deformation through crystallographic slip, slip-system hardening, and lattice rotation~\cite{roters2010overview,kalidindi1992crystallographic}. Combined with polycrystal homogenization, it links the initial crystallographic texture to the macroscopic response and to the texture evolution caused by deformation~\cite{lebensohn1993self,miehe1999computational,zecevic2018latent}. This link matters in deep drawing, where a material point follows a non-proportional loading path of stretching, compression, shear, and bending as it moves through the flange, over the die radius, and into the cup wall~\cite{tian2017earing,barrett2019deep}.

In a two-scale simulation, the finite-element (FE) model of a component obtains its material response from a microstructure model that is evaluated at each integration point, which gives the deformation of the component together with the evolution of its microstructure~\cite{miehe1999computational,feyel2000fe2,geers2010multiscale}. Self-consistent polycrystal plasticity models have been coupled with FE formulations to simulate sheet-metal forming, including the deep drawing of AA6022-T4 sheet~\cite{barrett2019deep,zecevic2017modeling}. Full-field approaches go beyond such mean-field homogenization and resolve the heterogeneous response of the polycrystal. In the FE--FFT method, a crystal-plasticity spectral solver based on the fast Fourier transform (FFT) is evaluated at the FE integration points, so that the macroscopic deformation and the microstructural response are computed concurrently~\cite{han2020multiscale}. Alternatively, FFT-based virtual experiments have been used to update a phenomenological anisotropic yield function during a deep-drawing simulation, which brings texture-dependent anisotropy into an inexpensive macroscopic model~\cite{han2020using}.

Polycrystal models nevertheless remain expensive in component-scale simulations, because the microscopic response must be evaluated at every integration point in every increment. This cost motivates reduced-order multiscale models that keep the microstructural physics at a fraction of the cost of each microscale evaluation.

Machine-learning surrogates have been proposed for this purpose. Bonatti et al.~\cite{bonatti2022cp} trained a recurrent neural network (RNN) on homogenized stress responses from full-field crystal-plasticity simulations. Liu et al.~\cite{liu2023learning} introduced a recurrent neural operator that captures history dependence through learned internal variables, although their three-dimensional applications relied on Taylor homogenization. Weeks and Stebner~\cite{weeks2025efficient} used RNN surrogates of viscoplastic self-consistent homogenization to predict the constitutive response and the texture evolution in incremental sheet forming. These surrogates reduce the cost considerably, but their accuracy depends on the loading histories in the training data and on the homogenization scheme that generated them. An alternative is a reduced-order model that keeps the crystal-plasticity law itself and approximates only the interactions between grains.

Deep material networks (DMNs) are such a model. A DMN represents the mechanical interactions within a heterogeneous microstructure by a trainable assembly of micromechanical homogenization building blocks~\cite{wei2026deep,liu2019deep,liu2019exploring,noels2022micromechanics}. Instead of learning stress--strain relationships, it identifies a reduced representation of the microstructural interactions from effective linear-elastic stiffness tensors computed by full-field simulations of a representative volume element (RVE)~\cite{gajek2020micromechanics,shin2023deep,srinivas2026rapid,wan2024decoding}. The constitutive laws of the constituents are kept at the material nodes of the trained network, so that nonlinear, history-dependent responses follow from the learned interaction parameters without further training for new loading paths. This separation of learned interactions from constitutive behavior has been used for nonlinear multiscale simulations of components~\cite{gajek2021fe,wei2023ls,srinivas2026deep}. The orientation-aware interaction-based deep material network (ODMN) extends DMNs to polycrystals. It assigns a crystallographic orientation to each material node and tracks its evolution during plastic deformation, which gives a texture-dependent anisotropic response~\cite{wei2025orientation,wei2025crystallographic}.

A recent extension of the ODMN~\cite{weiGSHmanuscript} prescribes the initial texture of the network through generalized spherical harmonics (GSH), following the compact texture representation of Marki and Knezevic~\cite{marki2024nonlinear}. Because the material nodes then carry the measured texture, the network represents texture-dependent anisotropy and the subsequent texture evolution more accurately at the same cost. The approach was demonstrated on RVEs of Inconel 625 and on a thermomechanically loaded L-bracket~\cite{weiGSHmanuscript}. The present study applies the GSH-initialized ODMN to sheet-metal forming, where directional plastic flow and texture evolution determine responses that can be compared with experiments.

This work develops FE--ODMN, a two-scale framework for crystal-plasticity simulation of the deep drawing of AA6022-T4 aluminum sheet. GSH texture compaction provides 16, 64, or 256 weighted orientations for the material nodes, and the interaction directions of the network are trained on homogenized linear-elastic stiffness tensors from full-field RVE simulations. The trained ODMN then acts as a reduced-order RVE at each integration point of an implicit finite-strain FE analysis. It returns the homogenized stress and tangent, so that no macroscopic constitutive model is needed, and it tracks the crystallographic texture at that point. At the RVE scale, the networks are verified against full-field spectral crystal-plasticity simulations for seven loading directions and for load reversals. At the component scale, the predicted punch force, wall-thickness distribution, and earing profile are compared with published measurements. The influence of the network depth on accuracy and cost is quantified at both scales, and the evolving material-node orientations give the local texture evolution in the drawn cup.

The paper is organized as follows. Section~\ref{sec:method} presents the FE--ODMN framework and the FE model of the cup-drawing test. Section~\ref{sec:material} describes the RVE built from the published texture, its compact textures, and the trained networks. Section~\ref{sec:rve} verifies the networks against full-field simulations of the same RVE. Section~\ref{sec:cup} reports the two-scale cup-drawing predictions, and Sections~\ref{sec:discussion} and~\ref{sec:conclusions} discuss and conclude the work.

\section{FE--ODMN framework}
\label{sec:method}

\begin{figure}[t]
  \centering
  \includegraphics[width=\textwidth]{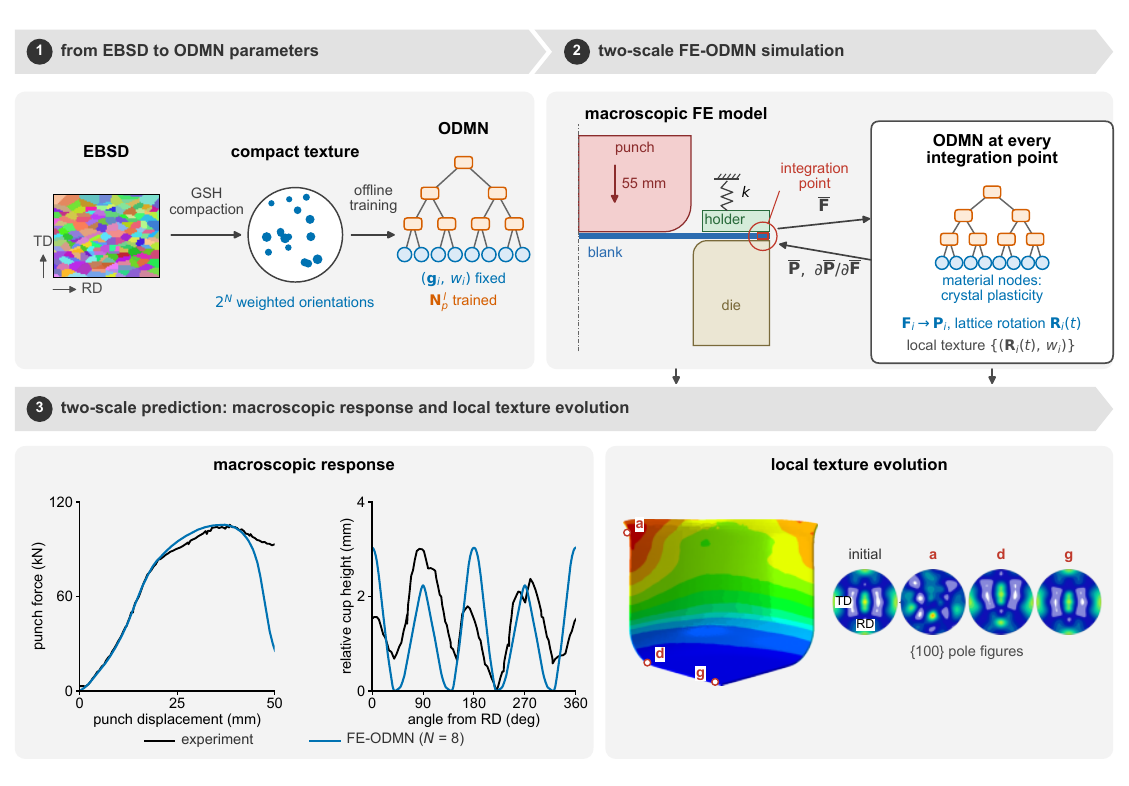}
  \caption{Overview of the FE--ODMN framework. (1)~The crystallographic texture characterized by electron backscatter diffraction (EBSD) is represented by $2^{\Nleaf}$ weighted orientations through GSH compaction. These orientations are assigned to the material nodes of the ODMN, and the interaction directions $\bm N_p^l$ are trained on linear-elastic homogenization data from full-field RVE simulations. (2)~The trained ODMN is evaluated as a reduced-order RVE at each integration point of the macroscopic FE simulation. Given the macroscopic deformation gradient $\Fb$, the network returns the first Piola--Kirchhoff stress $\Pb$ and the consistent tangent $\partial\Pb/\partial\Fb$, so that no macroscopic constitutive model is needed. (3)~Two-scale predictions of the punch force and the cup earing profile for $\Nleaf=8$, compared with the measurements reported by Barrett and Knezevic~\cite{barrett2019deep}. The EBSD map is schematic.}
  \label{fig:framework}
\end{figure}

FE--ODMN builds on the GSH-initialized ODMN of our previous study~\cite{weiGSHmanuscript}, specialized here to isothermal, slip-based crystal plasticity. The framework has three stages (Fig.~\ref{fig:framework}). First, GSH compaction represents the measured crystallographic texture by $N_m=2^N$ weighted orientations, where $N$ is the depth of the network and $N_m$ is the number of its material nodes, and each orientation is assigned to one node. Second, the interaction directions of the network are trained offline on effective linear-elastic stiffness tensors from full-field RVE simulations. Third, the trained ODMN is evaluated as a reduced-order RVE at each integration point of the macroscopic FE analysis, with finite-strain crystal plasticity at each material node.

The network parameters are trained once for the initial sheet texture and remain fixed during the forming simulation. Each FE integration point holds its own set of crystal-plasticity internal variables, including the current crystallographic orientations. The same network therefore gives a different response and a different texture evolution at each point of the sheet.

\subsection{Two-scale problem and the reduced-order RVE}
\label{sec:twoscale}

At each integration point of the macroscopic FE model, the trained ODMN takes the place of the full-field RVE. No macroscopic constitutive model is assumed. The macroscopic response is obtained directly from the RVE, and the only constitutive law in the framework is the single-crystal plasticity law at the material nodes. Given the macroscopic deformation gradient $\Fb$ and the internal states of the previous increment, the network returns the homogenized first Piola--Kirchhoff stress $\Pb$, the consistent algorithmic tangent $\overline{\mathbb L}=\partial\Pb/\partial\Fb$, and the updated crystal-plasticity states of its material nodes. Here and below, an overbar marks a macroscopic (homogenized) quantity. The macroscopic model handles the geometry, the boundary conditions, the tool contact, and global equilibrium, and the ODMN gives the homogenized RVE response by solving for the interactions among its material nodes. The two scales exchange $\Fb$, $\Pb$, and $\overline{\mathbb L}$, as in concurrent FE$^2$ methods~\cite{feyel2000fe2,geers2010multiscale}. In place of the RVE boundary-value problem of FE$^2$, the network solves a small equilibrium problem that satisfies the Hill--Mandel macrohomogeneity condition (Section~\ref{sec:odmn}). Each integration point stores the internal variables of all material nodes and the converged interaction vectors of the network, which initialize the solution of the next increment.

\subsection{Orientation-aware deep material network}
\label{sec:odmn}

An ODMN of depth $N$ has $N_m=2^N$ material nodes $\mathcal M_i$, $i=1,\ldots,N_m$. Each node represents a subdomain of the RVE with volume fraction $w_i$ and crystallographic orientation $\mathbf g_i$, where $\sum_i w_i=1$. The weighted orientation set $\{(\mathbf g_i,w_i)\}_{i=1}^{N_m}$ is a discrete representation of the orientation distribution function (ODF). The material nodes are homogenized through a binary tree of two-phase laminate building blocks, indexed by the level $l=0,\ldots,N-1$, with $l=0$ at the root, and by the position $p=0,\ldots,2^l-1$ within the level. Building block $(l,p)$ joins two child subdomains, which contain the material-node sets $\mathcal J^0_{l,p}$ and $\mathcal J^1_{l,p}$, across an interface with unit normal $\bm N_p^l$, called the interaction direction. Each interaction direction is parameterized by two dimensionless angular variables $(\theta_p^l,\phi_p^l)$ as $\bm N_p^l=[\cos(2\pi\phi_p^l)\sin(\pi\theta_p^l),\,\sin(2\pi\phi_p^l)\sin(\pi\theta_p^l),\,\cos(\pi\theta_p^l)]^{\mathsf T}$~\cite{weiGSHmanuscript}.

In the GSH-initialized ODMN, the orientations and weights of the material nodes are prescribed by the texture compaction of Section~\ref{sec:gsh} and stay fixed during training. The only trainable parameters are the $2(2^N-1)$ interaction angles, two for each of the $2^N-1$ building blocks. No constitutive parameter is fitted during training. The single-crystal constitutive model and its parameters are given in Appendix~\ref{sec:cp}.

The interaction angles are identified offline from linear-elastic homogenization data, following Wei et al.~\cite{weiGSHmanuscript}. Training sample $s$ pairs a randomly sampled cubic single-crystal stiffness $\bm C^{\mathrm c,(s)}$, which is assigned to all grains of the RVE and rotated to the orientation of each grain, with the effective stiffness $\overline{\bm C}^{\mathrm{DNS},(s)}$ of the RVE. The latter is computed by full-field direct numerical simulation (DNS) with the spectral solver of the D\"usseldorf Advanced Material Simulation Kit (DAMASK)~\cite{roters2019damask}. The ODMN prediction $\overline{\bm C}^{\mathrm{ODMN},(s)}$ follows from recursive homogenization of the material-node stiffnesses along the binary tree. The set of interaction angles $\boldsymbol\Theta_{\mathrm{GSH}}=\{\theta_p^l,\phi_p^l\}$ is optimized by minimizing the loss function $\mathcal L$, the mean relative error of the effective stiffness,
\begin{equation}
 \mathcal L(\boldsymbol\Theta_{\mathrm{GSH}})
 =\frac{1}{N_{\mathrm{batch}}}\sum_{s=1}^{N_{\mathrm{batch}}}
 \frac{\|\overline{\bm C}^{\mathrm{ODMN},(s)}
          -\overline{\bm C}^{\mathrm{DNS},(s)}\|_{\mathrm F}}
      {\|\overline{\bm C}^{\mathrm{DNS},(s)}\|_{\mathrm F}},
 \label{eq:training_loss}
\end{equation}
where $\|\cdot\|_{\mathrm F}$ is the Frobenius norm and $N_{\mathrm{batch}}$ is the minibatch size. The two-input homogenization operator of the building blocks, its recursive assembly, and the training procedure are described in Wei et al.~\cite{weiGSHmanuscript}.

In the online evaluation, the trained interaction directions are fixed, and one interaction vector $\bm a_p^l$ per building block is determined from the equilibrium of the network. The local deformation gradient of material node $\mathcal M_i$ is
\begin{equation}
 \bm F_i=\Fb+\sum_{l=0}^{N-1}\sum_{p=0}^{2^l-1}
 \alpha_{l,p}^{i}\,\bm a_p^l\otimes\bm N_p^l,
 \label{eq:localization}
\end{equation}
where $\otimes$ is the dyadic product and the localization coefficients are
\begin{equation}
 \alpha_{l,p}^{i}=
 \begin{cases}
 \displaystyle \frac{1}{W_{l,p}^0},
 & \mathcal M_i\in\mathcal J^0_{l,p},\\[6pt]
 \displaystyle -\frac{1}{W_{l,p}^1},
 & \mathcal M_i\in\mathcal J^1_{l,p},\\[6pt]
 0, & \text{otherwise},
 \end{cases}
 \qquad
 W_{l,p}^k=\sum_{j\in\mathcal J^k_{l,p}}w_j.
 \label{eq:localization_coeff}
\end{equation}

\noindent Here $W_{l,p}^k$ is the total weight of the material nodes in child subdomain $k\in\{0,1\}$ of building block $(l,p)$. For given interaction vectors, Eq.~\eqref{eq:localization} gives the local deformation gradients, and the crystal-plasticity update of Appendix~\ref{sec:cp} returns the first Piola--Kirchhoff stress $\bm P_i$ and the consistent tangent of each material node. The Hill--Mandel macrohomogeneity condition then requires the interface equilibrium residuals to vanish,
\begin{equation}
 \bm r_p^l=
 \sum_{i=1}^{N_m}w_i\,\alpha_{l,p}^{i}\,
 \bm P_i\cdot\bm N_p^l=\bm 0,
 \qquad l=0,\ldots,N-1,\quad p=0,\ldots,2^l-1.
 \label{eq:network_equilibrium}
\end{equation}
Each residual $\bm r_p^l$ is the traction imbalance between the two child subdomains of building block $(l,p)$. Together, the residuals form a system of $3(2^N-1)$ nonlinear equations for the interaction vectors.

With the interaction vectors collected in the vector $\mathbf A$, Eq.~\eqref{eq:localization} reads $\operatorname{vec}(\bm F_i)=\operatorname{vec}(\Fb)+\bm D_i\mathbf A$, where $\operatorname{vec}(\cdot)$ denotes row-wise vectorization and $\bm D_i$ is the localization matrix of node $i$, assembled from the coefficients $\alpha_{l,p}^{i}$ and the interaction directions. The assembled residual $\bm r$ and its Jacobian $\bm K$ are
\begin{equation}
 \bm r=\sum_{i=1}^{N_m}w_i\bm D_i^{\mathsf T}\operatorname{vec}(\bm P_i),
 \qquad
 \bm K=\sum_{i=1}^{N_m}w_i\bm D_i^{\mathsf T}\bm{\mathsf L}_i\bm D_i,
 \qquad
 \bm{\mathsf L}_i=\frac{\partial\operatorname{vec}(\bm P_i)}
                        {\partial\operatorname{vec}(\bm F_i)},
 \label{eq:network_newton}
\end{equation}
where $\bm{\mathsf L}_i$ is the consistent tangent of material node $i$ in matrix form. The interaction vectors are updated by Newton--Raphson iterations, $\bm K\,\delta\mathbf A=-\bm r$ with the increment $\delta\mathbf A$, combined with a backtracking line search. The local deformation gradients, the stresses, and the residuals are re-evaluated after each update until convergence.

At convergence, the homogenized first Piola--Kirchhoff stress is the weighted average of the nodal stresses. The consistent macroscopic tangent, written as the matrix $\Lb$ that represents $\overline{\mathbb L}$, follows from implicit differentiation of the network equilibrium equations,
\begin{align}
 \Pb&=\sum_{i=1}^{N_m}w_i\bm P_i,\label{eq:homogenized_stress}\\
 \Lb&=\sum_{i=1}^{N_m}w_i\bm{\mathsf L}_i
 -\Big(\sum_{i=1}^{N_m}w_i\bm{\mathsf L}_i\bm D_i\Big)\bm K^{-1}
  \Big(\sum_{j=1}^{N_m}w_j\bm D_j^{\mathsf T}\bm{\mathsf L}_j\Big).
 \label{eq:condensed_tangent}
\end{align}

The ODMN is coupled to Abaqus/Standard through the user material subroutine (UMAT) interface. The Cauchy stress and the consistent Jaumann-rate tangent required by Abaqus are derived from $\Pb$ and $\Lb$, respectively. If the network iteration fails to converge, the increment is first subdivided locally, and a cutback of the global increment is requested only if the substeps also fail. Implementation details are given in Wei et al.~\cite{weiGSHmanuscript}.

\subsection{GSH texture compaction and material nodes}
\label{sec:gsh}

The crystallographic texture enters the ODMN through its material nodes. The RVE texture is a set of weighted grain orientations $\{(\mathbf g_q^{\mathrm{RVE}},v_q)\}_{q=1}^{N_{\mathrm{RVE}}}$, where $N_{\mathrm{RVE}}$ is the number of grains, $\mathbf g_q^{\mathrm{RVE}}$ is the orientation of grain $q$, and $v_q$ is its volume fraction, with $\sum_q v_q=1$. Its ODF is characterized by the normalized GSH coefficients
\begin{equation}
 \widehat C_l^{\mu\nu,\mathrm{RVE}}
  =\sum_{q=1}^{N_{\mathrm{RVE}}}v_q\,[\tridotT(\mathbf g_q^{\mathrm{RVE}})]^*,
 \qquad 0\le l\le L,\quad l\ \mathrm{even},
 \label{eq:gsh_coefficients}
\end{equation}
where $\tridotT$ are the symmetrized GSH basis functions for cubic crystal symmetry and triclinic sample symmetry, $l$ is the harmonic degree, $\mu$ and $\nu$ number the independent basis functions of degree $l$, $(\cdot)^*$ denotes complex conjugation, and $L$ is the truncation order. In this subsection, $l$ is the harmonic degree, as in the texture literature, and not the tree level of Section~\ref{sec:odmn}.

Following Marki and Knezevic~\cite{marki2024nonlinear}, the RVE texture is compacted to the $N_m=2^N$ weighted material-node orientations $\{(\mathbf g_i,w_i)\}_{i=1}^{N_m}$ by matching the GSH coefficients up to order $L$. Both the orientations and the weights are optimized, so that far fewer orientations than grains reproduce the texture. The truncation order $L$ is chosen for each network depth $N$ in view of the $4N_m-1$ degrees of freedom of the compact texture, namely three orientation parameters and one weight per node, less the constraint on the sum of the weights.

The accuracy of the compaction, and later the amount of texture evolution, is measured by the texture-difference index~\cite{li2005texture}
\begin{equation}
 \Td=\frac{\|f-f_{\mathrm{ref}}\|^{2}}{\|f_{\mathrm{ref}}\|^{2}},
 \label{eq:Td}
\end{equation}
where $f$ is the ODF under consideration, $f_{\mathrm{ref}}$ is the reference ODF, and $\|\cdot\|$ is the $L^2$ norm over orientation space, so that $\Td=0$ for identical textures. Both ODFs are reconstructed from the weighted orientations with a de la Vall\'ee Poussin kernel of $10^\circ$ halfwidth in MTEX~\cite{bachmann2010texture}. All orientations refer to the sample frame of the sheet, with $x$ along the rolling direction (RD), $y$ along the normal direction (ND), and $z$ along the transverse direction (TD).

\subsection{Finite-element model of the cup-drawing test}
\label{sec:fe_model}

\begin{figure}[t]
  \centering
  \includegraphics[width=0.95\textwidth]{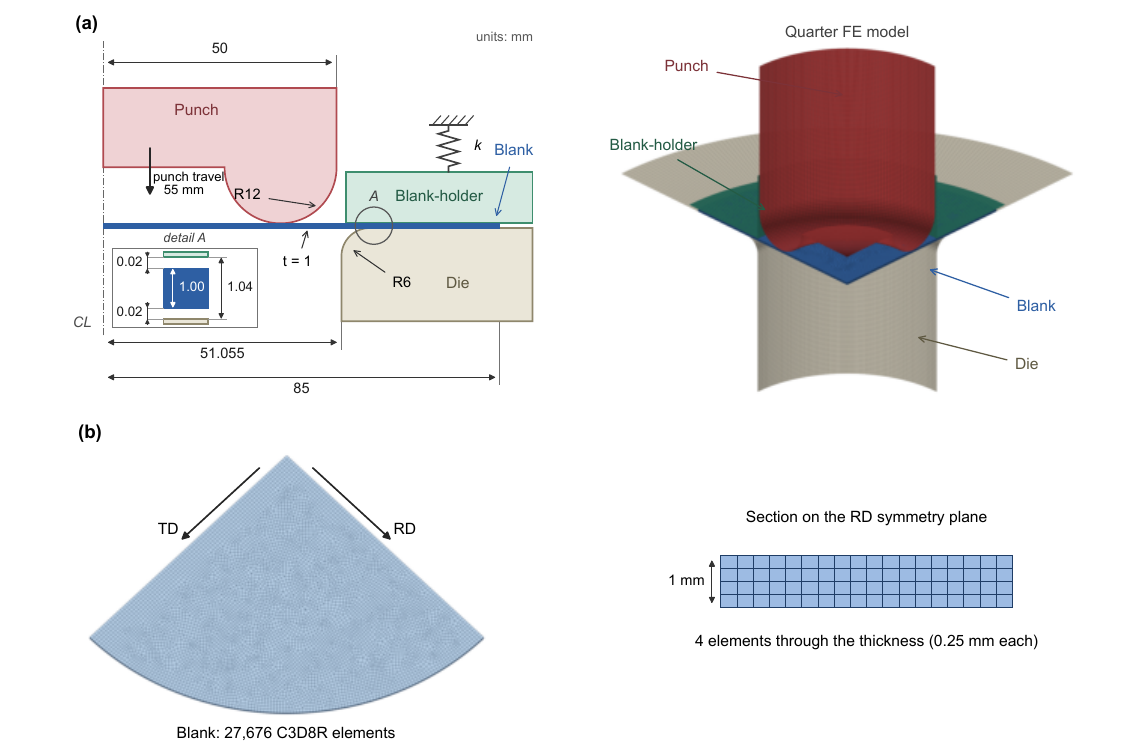}
  \caption{FE model of the cup-drawing test. (a)~Tooling geometry and quarter model, with an enlarged view (detail~A) of the clearances between the blank, the blank holder, and the die. (b)~Blank mesh with four elements through the thickness. CL, RD, and TD denote the centerline, the rolling direction, and the transverse direction.}
  \label{fig:setup}
\end{figure}

The FE model follows the cylindrical cup-drawing experiment of Tian et al.~\cite{tian2017earing}, in which lubricated blanks of 170~mm diameter and 1~mm thickness were drawn under a spacer-controlled blank holder. The tooling geometry is shown in Fig.~\ref{fig:setup}, and the model parameters are listed in Table~\ref{tab:fe}. A quarter of the blank is discretized with 27\,676 eight-node brick elements with reduced integration (C3D8R in Abaqus) and enhanced hourglass control, with four elements through the thickness and an in-plane element size of about 0.9~mm. This in-plane size is finer than the radial element size of 1.25~mm that Barrett and Knezevic~\cite{barrett2019deep} found sufficient for a mesh-converged solution of the same problem. Symmetry boundary conditions are imposed on the planes normal to RD and TD.

The punch, the die, and the blank holder are discrete rigid surfaces. The die is fixed, and the punch travels 55~mm along ND with a smooth displacement history. The blank holder is free to translate along ND and is connected to ground by a spring of stiffness $1.25\times10^{6}$~N/mm, which represents the compliance of the spacer-controlled blank-holder system as in Barrett and Knezevic~\cite{barrett2019deep}.

Surface-to-surface contact is defined between the sheet and each tool. Tangential contact follows Coulomb friction with the friction coefficient $\mu$ in a penalty formulation, and normal contact uses the exponential pressure--overclosure relationship
\begin{equation}
 p=\begin{cases}
 0, & h\le -c_0,\\
 \dfrac{p_0}{\mathrm e-1}\Big(\dfrac{h}{c_0}+1\Big)
 \Big[\exp\Big(\dfrac{h}{c_0}+1\Big)-1\Big], & h>-c_0,
 \end{cases}
 \label{eq:contact}
\end{equation}
where $p$ is the contact pressure, $h$ is the overclosure (negative for a clearance), $c_0$ is the clearance at which the contact pressure starts to build up, $p_0$ is the pressure at zero overclosure, and $\mathrm e$ is the base of the natural logarithm. The sheet--die and sheet--holder interfaces use $c_0=0.01$~mm, $p_0=200$~MPa, and $\mu=0.075$. The sheet--punch interface uses a softer law with $c_0=0.02$~mm, $p_0=50$~MPa, and $\mu=0.30$.

The stiffer contact law at the die and the holder retains part of the late-stage punch force that arises when the wrinkled rim is ironed through the die throat, a force that a softer contact law suppresses~\cite{barrett2019deep}. The higher friction coefficient at the punch limits thinning around the punch nose. The contact parameters were calibrated against the measured peak punch force and minimum wall thickness with the $N=6$ network and then held fixed, so that only the network differs among the three cup-drawing simulations. Drawing, tool removal, and springback are computed in three successive static steps with automatic stabilization.

\begin{table}[t]
  \centering
  \caption{Parameters of the cup-drawing model.}
  \label{tab:fe}
  \footnotesize
  \setlength{\tabcolsep}{4pt}
  \begin{tabular}{@{}ll@{}}
    \toprule
    Parameter & Specification \\
    \midrule
    Blank & Diameter 170~mm; thickness 1~mm \\
    Punch & Diameter 100~mm; profile radius 12~mm; stroke 55~mm \\
    Die & Opening diameter 102.11~mm; profile radius 6~mm \\
    Blank holder & Spring stiffness $1.25\times10^{6}$~N/mm; clearance 1.04~mm \\
    Blank mesh & Quarter model; 27\,676 C3D8R elements; four through thickness \\
    Tools & Discrete rigid surfaces \\
    Die and holder contact & Exponential law; $c_0=0.01$~mm; $p_0=200$~MPa; $\mu=0.075$ \\
    Punch contact & Exponential law; $c_0=0.02$~mm; $p_0=50$~MPa; $\mu=0.30$ \\
    Analysis sequence & Drawing, tool removal, and springback \\
    Solution procedure & Static with automatic stabilization \\
    \bottomrule
  \end{tabular}
\end{table}

\section{RVE construction and network training}
\label{sec:material}

\subsection{Sheet texture and representative volume element}
\label{sec:rve_model}

The material is a 1~mm thick sheet of the automotive aluminum alloy AA6022-T4. Its texture, mechanical response, and cup-drawing behavior were characterized by Tian et al.~\cite{tian2017earing} and Barrett and Knezevic~\cite{barrett2019deep}. All experimental data used here are taken from these two studies, and the measured curves were digitized from the published figures. Because the raw electron backscatter diffraction (EBSD) data are not available, a set of orientations was generated to reproduce the published EBSD texture and serves as the texture input of the framework.

\begin{figure}[t]
  \centering
  \includegraphics[width=0.55\textwidth]{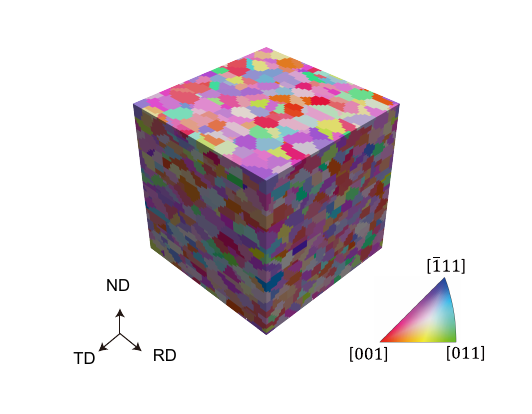}
  \caption{RVE with 2\,000 grains built from the published texture. The grains are colored by the inverse pole figure of RD.}
  \label{fig:rve}
\end{figure}

The RVE in Fig.~\ref{fig:rve} contains 2\,000 grains on a periodic grid of $64\times64\times64$ voxels, with orientations taken from the generated set. The grains are elongated along RD, with mean chord lengths along RD, TD, and ND in the ratio $1.66:1.27:1$. They are less elongated than the grains reported by Barrett and Knezevic~\cite{barrett2019deep}, who estimated axis lengths of about 50, 34, and \qty{20}{\micro\metre}, a ratio of $2.5:1.7:1$. The pole figures of the RVE are shown in Fig.~\ref{fig:pf}(a). The RVE texture is compacted onto the material nodes, and the same RVE provides the full-field solutions for training (Section~\ref{sec:odmn}) and for verification (Section~\ref{sec:rve}).

\subsection{Compact textures and offline training}
\label{sec:training_results}

\begin{figure}[t]
  \centering
  \includegraphics[width=0.8\textwidth]{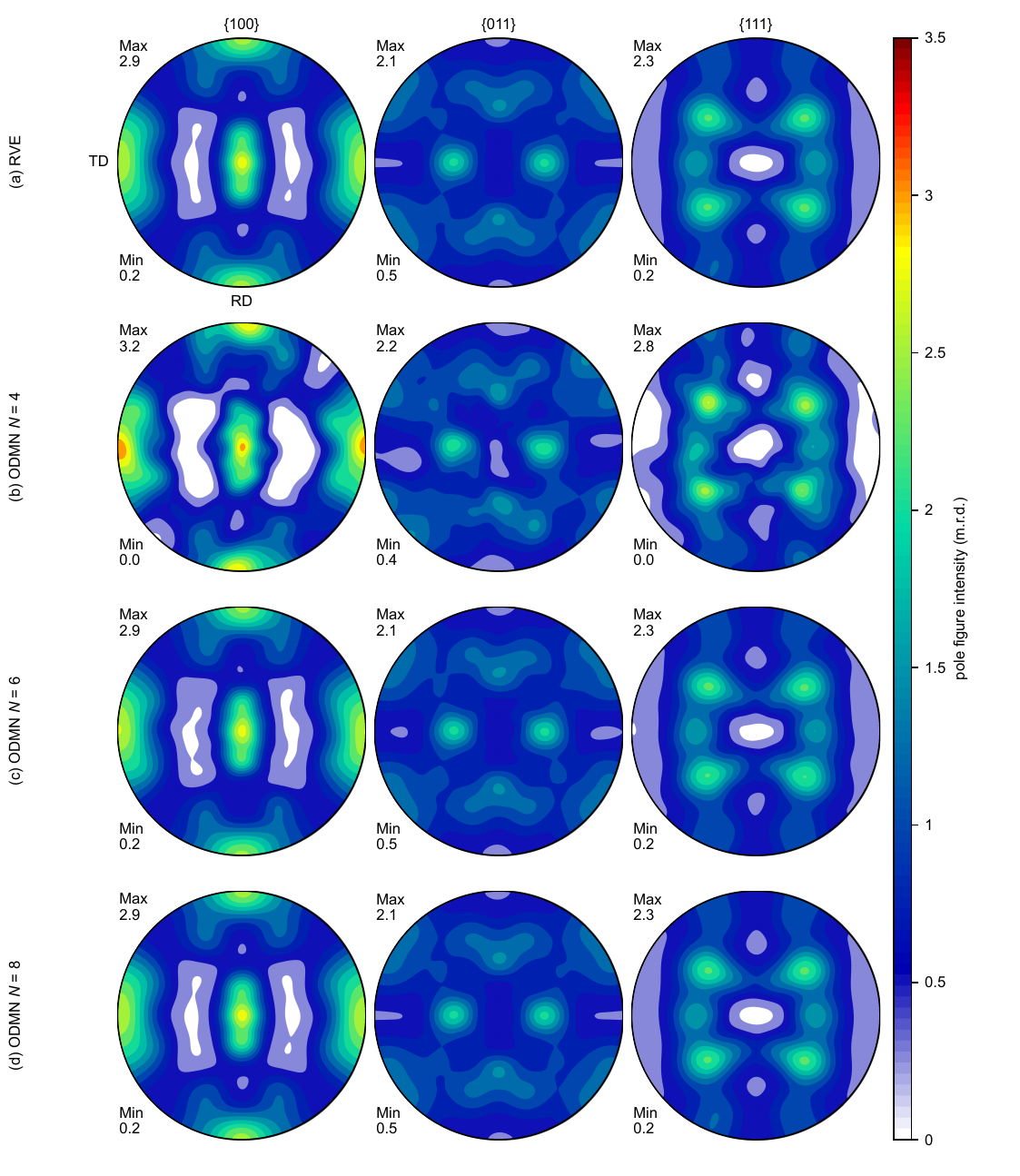}
  \caption{$\{100\}$, $\{011\}$, and $\{111\}$ pole figures of (a) the RVE of Fig.~\ref{fig:rve} and (b--d) the compact textures of the $N=4$, 6, and 8 networks. Stereographic projection, with intensities in multiples of a random distribution (m.r.d.).}
  \label{fig:pf}
\end{figure}

Three network depths are considered, $N=4$, 6, and 8, with 16, 64, and 256 material nodes. The GSH truncation orders are $L=12$, 14, and 24, as adopted by Wei et al.~\cite{weiGSHmanuscript}. The texture-difference index with respect to the 2\,000-grain RVE falls by about one order of magnitude with each step in depth (Table~\ref{tab:gsh}), in line with the findings of Wei et al.~\cite{weiGSHmanuscript}. The compact textures with 64 and 256 orientations reproduce the pole figures of the RVE (Fig.~\ref{fig:pf}). The 16-orientation texture is less accurate because it has too few degrees of freedom. Its $\{100\}$ and $\{111\}$ pole-figure maxima are 3.2 and 2.8 multiples of a random distribution (m.r.d.), against 2.9 and 2.3 for the RVE.

\begin{table}[t]
  \centering
  \caption{GSH compaction and offline training of the three networks. $N_m$ is the number of material nodes, $L$ is the GSH truncation order, $\Td$ is taken with respect to the RVE texture, the pole-figure maxima are in m.r.d., $n_\Theta$ is the number of trained interaction angles, and the validation error is the lowest value of Eq.~\eqref{eq:training_loss} on the validation set.}
  \label{tab:gsh}
  \footnotesize
  \setlength{\tabcolsep}{2.5pt}
  \begin{tabular}{lcccccc}
    \toprule
    & $N_m$ & $L$ & $\Td$ & $\{100\}$ / $\{011\}$ / $\{111\}$ max. & $n_\Theta$ & validation error \\
    \midrule
    RVE (2\,000 grains) & --  & --  & --  & 2.9 / 2.1 / 2.3 & -- & -- \\
    $\Nleaf$ = 4      & 16  & 12  & 0.177   & 3.2 / 2.2 / 2.8 & 30  & $8.7\times10^{-3}$ \\
    $\Nleaf$ = 6      & 64  & 14  & 0.0172  & 2.9 / 2.1 / 2.3 & 126 & $2.9\times10^{-4}$ \\
    $\Nleaf$ = 8      & 256 & 24  & 0.00245 & 2.9 / 2.1 / 2.3 & 510 & $9.8\times10^{-5}$ \\
    \bottomrule
  \end{tabular}
\end{table}

\begin{figure}[t]
  \centering
  \includegraphics[width=\textwidth]{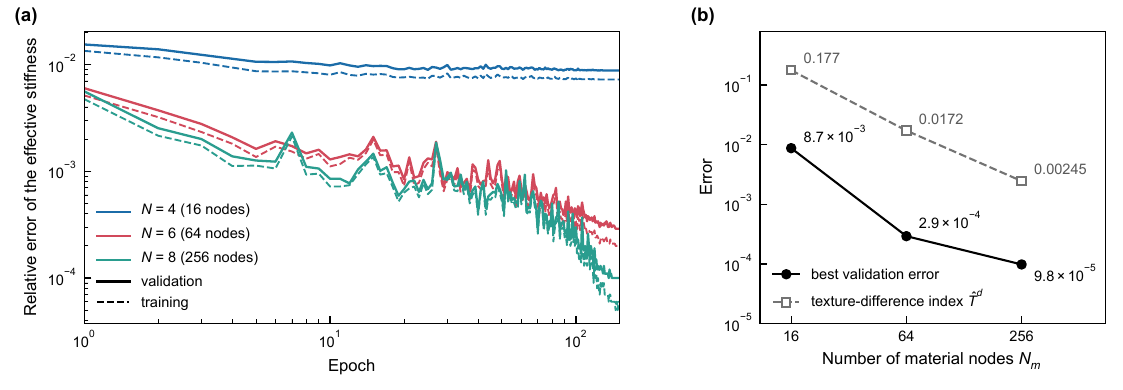}
  \caption{Offline training of the three networks. (a)~Training (dashed) and validation (solid) errors of the effective stiffness, Eq.~\eqref{eq:training_loss}, against the epoch. (b)~Lowest validation error of each network and texture-difference index $\Td$ of its compact texture against the number of material nodes.}
  \label{fig:training}
\end{figure}

The interaction directions of the three networks were trained for 150 epochs on the same linear-elastic dataset of 500 samples, 400 for training and 100 for validation. The error of the $N=4$ network levels off at about $9\times10^{-3}$ after 20 epochs (Fig.~\ref{fig:training}(a)), which points to the capacity of the network, not the optimization, as the limit. The errors of the $N=6$ and 8 networks keep decreasing over the last 100 epochs. The lowest validation errors are $8.7\times10^{-3}$, $2.9\times10^{-4}$, and $9.8\times10^{-5}$ for $N=4$, 6, and 8 (Table~\ref{tab:gsh}), and they decrease monotonically with the number of material nodes (Fig.~\ref{fig:training}(b)).

\section{RVE-level verification}
\label{sec:rve}

The networks are verified at the RVE level before they are used at the component scale. The full-field RVE solutions are first compared with the measured stress--strain curves and Lankford coefficients, which checks the single-crystal constitutive law and establishes the RVE as the reference. The Lankford coefficient $\rvalue$ is the ratio of the plastic strain in the width direction to that in the thickness direction of a uniaxial tensile specimen. The ODMN predictions are then compared with the full-field solutions for the same loading paths.

The full-field reference solutions are computed for the RVE with the spectral solver of DAMASK~\cite{roters2019damask,eisenlohr2013spectral,shanthraj2015numerically} and the crystal-plasticity law of Appendix~\ref{sec:cp}. They are labeled DAMASK-FFT in the figures and tables. Eleven loading paths are used at every stage of the verification. Seven are uniaxial tension tests at 0, 15, 30, 45, 60, 75, and 90$^\circ$ to RD, up to a stretch of 1.22 at a stretch rate of $10^{-3}$~s$^{-1}$. The stress state is uniaxial, so that the width and the thickness contract freely and the Lankford coefficient can be evaluated. Four are tension--compression--tension cycles along RD at the four tensile pre-strains of the measurements in Section~\ref{sec:rve_vs_exp}. The spectral solver imposes uniaxial stress through mixed boundary conditions. Each network is evaluated along the same eleven paths with one C3D8R element in Abaqus/Standard, with uniaxial stress enforced by kinematic constraints on the element faces and with the displacement history of the spectral solver. The only difference between the two sets of solutions is that 16, 64, or 256 material nodes replace the $64\times64\times64$ grid.

\subsection{Full-field reference against the measurements}
\label{sec:rve_vs_exp}

\begin{figure}[t]
  \centering
  \includegraphics[width=\textwidth]{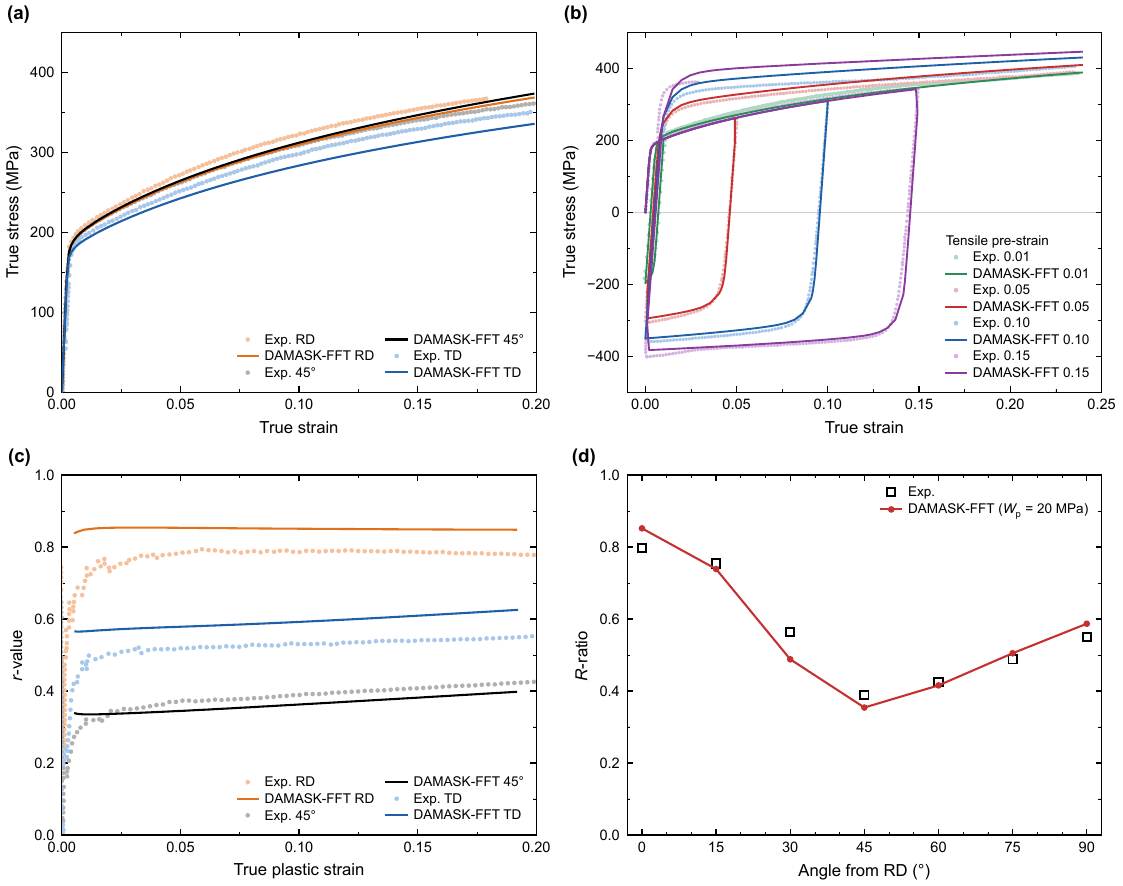}
  \caption{Full-field RVE simulations (DAMASK-FFT, solid lines) against the measurements reported by Barrett and Knezevic~\cite{barrett2019deep} (symbols). (a)~Uniaxial tension along RD, 45$^\circ$, and TD. (b)~Tension--compression--tension along RD at tensile pre-strains of about 0.01, 0.05, 0.10, and 0.15. (c)~Lankford coefficient against plastic strain. (d)~Lankford coefficient $\rvalue(\theta)$ against the angle $\theta$ from RD at a plastic work per unit volume $W_{\mathrm p}$ of 20~MPa. The axis labels $r$-value and $R$-ratio both denote the Lankford coefficient.}
  \label{fig:rve_vs_exp}
\end{figure}

Figure~\ref{fig:rve_vs_exp} compares the full-field RVE solutions with the measurements of Tian et al.~\cite{tian2017earing} and Barrett and Knezevic~\cite{barrett2019deep}. The measurements comprise uniaxial tension along RD, 45$^\circ$, and TD, tension--compression--tension tests along RD at four tensile pre-strains, and Lankford coefficients at seven in-plane angles.

The computed tensile flow stresses agree with the measurements within 5\,\% along RD, 45$^\circ$, and TD (Fig.~\ref{fig:rve_vs_exp}(a)). The RVE reproduces the lower flow stress along TD, but it places the flow stress at 45$^\circ$ slightly above that along RD, which is the reverse of the measured order. Under cyclic loading, the computed stress levels agree with the measurements within 6\,\% (Fig.~\ref{fig:rve_vs_exp}(b)). Both the computed and the measured Lankford coefficients change little with plastic strain beyond the first few percent, and the computed values along RD and TD lie somewhat above the measured ones (Fig.~\ref{fig:rve_vs_exp}(c)). The angular variation of the Lankford coefficient follows the measurements, including the minimum at 45$^\circ$ (Fig.~\ref{fig:rve_vs_exp}(d)). With $\rvalue_\theta$ denoting the Lankford coefficient at an angle $\theta$ from RD, the planar anisotropy parameter $\Delta\rvalue=(\rvalue_0+\rvalue_{90})/2-\rvalue_{45}$ is overestimated, at 0.37 against the measured 0.28.

\subsection{ODMN against the full-field simulations}
\label{sec:odmn_vs_dns}

\begin{figure}[t]
  \centering
  \includegraphics[width=\textwidth]{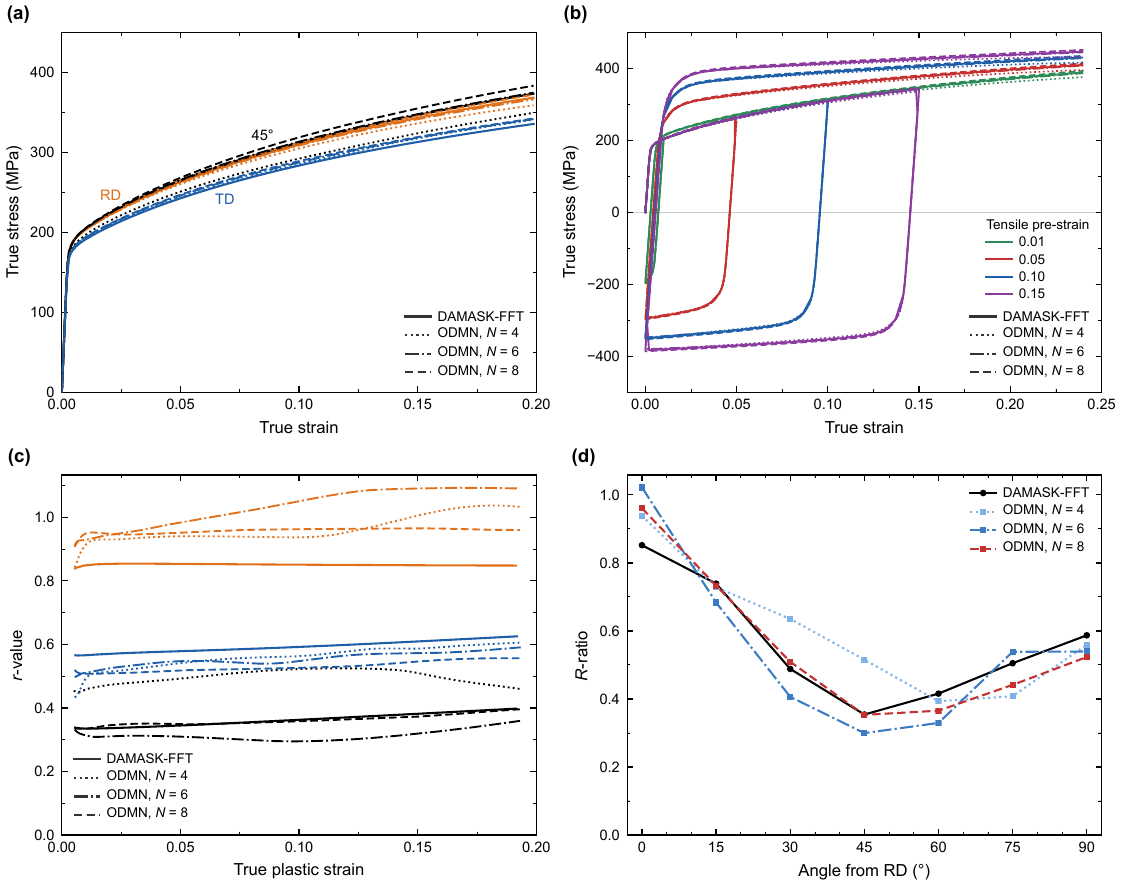}
  \caption{ODMN predictions with $N=4$ (dotted), 6 (dash-dotted), and 8 (dashed) against the full-field RVE simulations (DAMASK-FFT, solid lines). (a)~Uniaxial tension along RD, 45$^\circ$, and TD. (b)~Tension--compression--tension along RD at four tensile pre-strains. (c)~Lankford coefficient against plastic strain. (d)~Lankford coefficient $\rvalue(\theta)$ against the angle $\theta$ from RD at a plastic work per unit volume of 20~MPa. The axis labels $r$-value and $R$-ratio both denote the Lankford coefficient.}
  \label{fig:odmn_vs_fft}
\end{figure}

The $N=6$ and 8 networks reproduce the full-field stress--strain responses under uniaxial tension and under cyclic loading, and the $N=4$ network deviates more in both cases (Figs.~\ref{fig:odmn_vs_fft}(a) and (b)). The $N=6$ network has the lowest stress errors, and the $N=8$ network gives the most accurate Lankford coefficients. Table~\ref{tab:rve_level} lists the errors and the central processing unit (CPU) times.

\begin{table}[t]
  \centering
  \caption{Accuracy and computing time of the ODMN relative to the full-field RVE simulations (DAMASK-FFT). The errors are taken with respect to the full-field solution, the planar anisotropy parameter $\Delta\rvalue$ is listed as a value, and the CPU times are averages over the loading paths.}
  \label{tab:rve_level}
  \footnotesize
  \setlength{\tabcolsep}{4pt}
  \begin{tabular}{lcccc}
    \toprule
    & DAMASK-FFT & $N=4$ & $N=6$ & $N=8$ \\
    \midrule
    Tensile flow stress, maximum error (\%)
        & -- & 6.5 & 2.4 & 3.0 \\
    $\rvalue(\theta)$, mean absolute error
        & -- & 0.079 & 0.076 & 0.045 \\
    Planar anisotropy $\Delta\rvalue$
        & +0.37 & +0.23 & +0.48 & +0.39 \\
    Cyclic stress, maximum deviation (MPa)
        & -- & 14 & 3 & 5 \\
    CPU time per loading path (s)
        & 16\,660 & 2.8 & 4.5 & 16.9 \\
    Speed-up
        & 1 & 6\,000 & 3\,700 & 990 \\
    \bottomrule
  \end{tabular}
\end{table}

The Lankford coefficients depend more strongly on the network depth (Fig.~\ref{fig:odmn_vs_fft}(d)). Their mean absolute errors with respect to the full-field solution are 0.079, 0.076, and 0.045 for $N=4$, 6, and 8. The $N=6$ and 8 networks reproduce the minimum at 45$^\circ$, whereas the $N=4$ network places it at 60$^\circ$. The $N=6$ and 8 networks overestimate $\rvalue_0$ by 0.17 and 0.11, respectively, which adds to the discrepancy with the experiments, because the full-field RVE already overestimates $\rvalue_0$. For $N=6$, the Lankford coefficient along RD also drifts upward with plastic strain, whereas the full-field value stays constant (Fig.~\ref{fig:odmn_vs_fft}(c)).

The larger errors of the $N=4$ network may stem in part from its coarse texture, which overestimates the $\{100\}$ and $\{111\}$ pole-figure maxima (Fig.~\ref{fig:pf}). A deeper network represents the texture better and approximates the full-field response more accurately.

\section{Two-scale simulation of cup drawing}
\label{sec:cup}

\subsection{Punch force, wall thickness, and earing}
\label{sec:cup_macro}

\begin{figure}[t]
  \centering
  \includegraphics[width=\textwidth]{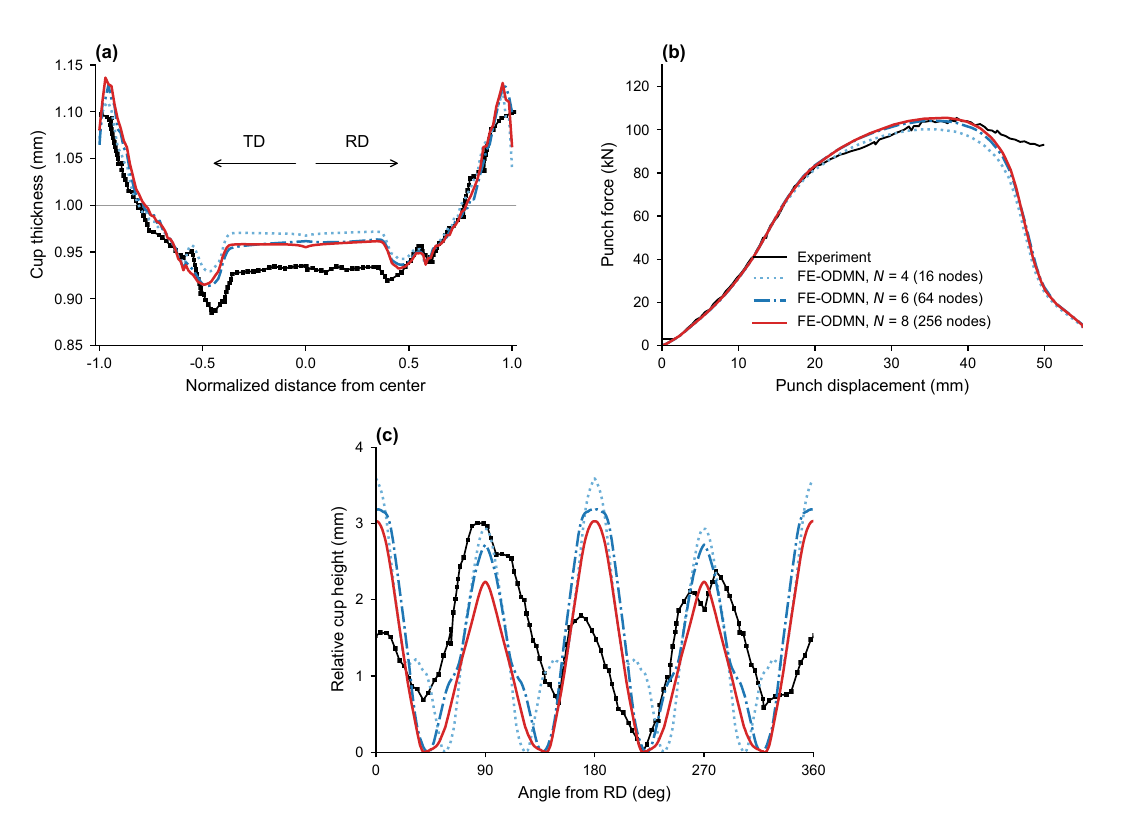}
    \caption{Cup-drawing predictions of FE--ODMN with $N=4$, 6, and 8 against the measurements reported by Barrett and Knezevic~\cite{barrett2019deep}. (a)~Wall thickness against the normalized distance from the cup center to the rim on the TD (left) and RD (right) sides. (b)~Punch force against punch displacement, with the legend for all panels. (c)~Cup-height profile after springback, relative to its lowest point, with the quarter-model predictions repeated over the full circumference.}
  \label{fig:cup_results}
\end{figure}

The $N=6$ and 8 models reproduce the measured punch-force curve, the trend of the wall thickness, and the four-ear pattern (Fig.~\ref{fig:cup_results}). Their peak-force errors are 1.2\,\% and 0.1\,\%. Their errors in the cup-height range, defined as the difference between the highest and the lowest point of the rim, are 5.9\,\% and 0.7\,\%. Both give the minimum wall thickness along TD within 3.4\,\% of the measurement. The $N=4$ model is less accurate, with errors of 5.0\,\% in the peak force and in the minimum wall thickness along TD, and of 19\,\% in the cup-height range. Table~\ref{tab:nstudy} summarizes the comparison.

The predicted wall-thickness profiles show the thinning near the punch profile radius and the thickening toward the rim (Fig.~\ref{fig:cup_results}(a)), although the predicted thickness is mostly above the measured one. The $N=6$ and 8 models give nearly the same minimum thickness along TD, 0.914 and 0.915~mm, which is closer to the measured 0.885~mm than the 0.929~mm of the $N=4$ model. The $N=8$ model places the minimum at a normalized distance of 0.49 from the cup center, against 0.46 in the experiment, and predicts less thinning along RD, where the minimum is 0.932~mm. Toward the rim, the thickness reaches 1.13 to 1.14~mm, slightly above the measured maximum of 1.10~mm, and then decreases at the free edge. The cup bottom is overpredicted in a similar way, by about 0.03~mm (3\,\%) for $N=6$ and 8.

The predicted punch forces follow the measured curve during the initial rise (Fig.~\ref{fig:cup_results}(b)). The $N=6$ and 8 models reach peak forces of 104.2 and 105.6~kN, close to the measured 105.5~kN, although both peaks occur slightly early, at strokes of 35.8 and 36.8~mm against the measured 38.5~mm. The $N=4$ model underpredicts the peak at 100.2~kN. The deviations are larger late in the process. The predicted force drops when the flange leaves the blank holder at a stroke of about 47~mm, whereas the measured force stays above 90~kN up to 50~mm, where the digitized curve ends. This difference may be attributed to ironing of the thickened and wrinkled rim as it passes through the die opening, where the radial clearance between the punch and the die is 1.055~mm and the measured wall thickness reaches 1.10~mm. The simulations capture this effect only in part, even with the stiffer contact law at the die.

After springback, the cup has four ears along RD and TD, with valleys at 45$^\circ$ to these directions (Fig.~\ref{fig:cup_results}(c)). These positions are consistent with the angular variation of the Lankford coefficient in Fig.~\ref{fig:rve_vs_exp}(d), which has its minimum at 45$^\circ$.

The $N=8$ model predicts a cup-height range of 3.03~mm, close to the measured 3.01~mm. The height of each ear is measured from the deeper of its two adjacent valleys. The predicted TD ear height is 2.24~mm, close to the local peak-to-valley height of about 2.4~mm estimated from the experimental profile. The predicted RD ear height, however, is 3.03~mm, whereas the two measured RD ears are 1.0 and 1.8~mm high. The overall range is reproduced, but the model predicts higher ears along RD than along TD, which reverses the measured order. This discrepancy is discussed in Section~\ref{sec:discussion}. The measured profile also departs from orthotropic symmetry. The quarter model forces opposite ears to have equal heights, although the RD and TD heights may differ, and cannot reproduce this asymmetry.

\begin{table}[t]
  \centering
    \caption{Cup-drawing predictions of the three networks against the measurements reported by Barrett and Knezevic~\cite{barrett2019deep}, with relative errors in parentheses. The ear heights are measured from the deeper adjacent valley, and the cup-height range is the difference between the highest and the lowest point of the rim.}
  \label{tab:nstudy}
  \scriptsize
  \setlength{\tabcolsep}{2pt}
  \begin{tabular}{lcccccc}
    \toprule
    & peak force & at stroke & min.\ TD wall & RD ear & TD ear & cup-height \\
    (material nodes) & (kN) & (mm) & thickness (mm) & height (mm) & height (mm) & range (mm) \\
    \midrule
    Experiment            & 105.5 & 38.5 & 0.885 & 1.0--1.8 & 2.4 & 3.01 \\
    $\Nleaf$ = 4 (16)     & 100.2 ($-$5.0\,\%) & 35.8 & 0.929 (+5.0\,\%) & 3.59 & 2.93 & 3.59 (+19\,\%) \\
    $\Nleaf$ = 6 (64)     & 104.2 ($-$1.2\,\%) & 35.8 & 0.914 (+3.3\,\%) & 3.19 & 2.72 & 3.19 (+5.9\,\%) \\
    $\Nleaf$ = 8 (256)    & 105.6 (+0.1\,\%)   & 36.8 & 0.915 (+3.4\,\%) & 3.03 & 2.24 & 3.03 (+0.7\,\%) \\
    \bottomrule
  \end{tabular}
\end{table}

\subsection{Convergence with network depth}
\label{sec:nstudy}

The same FE model and contact parameters are used for $N=4$, 6, and 8. With $N=6$, the network used to calibrate the contact parameters, the peak punch force and the minimum wall thickness are within 1.2\,\% and 3.3\,\% of the measurements (Table~\ref{tab:nstudy}). Increasing the depth to $N=8$ improves the cup-height range and the individual ear heights at four times the core-hours of $N=6$ (Section~\ref{sec:cost}). The peak-force error changes from $-$5.0\,\% for $N=4$ to $-$1.2\,\% for $N=6$ and +0.1\,\% for $N=8$, and the error in the cup-height range falls from +19\,\% to +5.9\,\% and +0.7\,\%.

The RVE-level results in Table~\ref{tab:rve_level} help to explain these differences. Earing depends on the directional flow stresses and on the angular variation of the Lankford coefficient. The $N=4$ network underpredicts the full-field flow stress at 45$^\circ$ by 6.5\,\% and places the minimum of the Lankford coefficient at 60$^\circ$ instead of 45$^\circ$. It also gives the lowest peak punch force and the highest ears of the three networks, and both errors may contribute.

The $N=6$ and 8 networks are closer to the full-field response. In the analytical earing model of Yoon et al.~\cite{yoon2006prediction}, the cup height at an angle $\theta$ from RD increases with $\rvalue_{\theta+90^\circ}/(1+\rvalue_{\theta+90^\circ})$ for a given tooling geometry, so that the height of the TD ears is governed by $\rvalue_0$. The larger overprediction of $\rvalue_0$ by the $N=6$ network, 0.17 against 0.11 for $N=8$, may therefore contribute to its higher TD ears. The TD ear of the $N=6$ model is 0.48~mm higher than that of the $N=8$ model, whereas the RD ears differ by only 0.16~mm. For $N=8$, the flow stresses are within 3\,\% of the full-field solution, and the mean absolute error of the Lankford coefficients is 0.045, against 0.076 for $N=6$. The $N=8$ model is also closest to the measured ear heights, although its RD ears remain too high.

The minimum wall thickness changes little beyond $N=6$. Its error falls from +5.0\,\% for $N=4$ to +3.3\,\% for $N=6$ and stays at +3.4\,\% for $N=8$. A deeper network alone is therefore unlikely to remove the remaining discrepancy, which may come from assumptions shared by the three models, namely the RVE, the crystal-plasticity law, and the description of contact and friction.

\subsection{Texture evolution in the drawn cup}
\label{sec:cup_texture}

\begin{figure}[t]
  \centering
  \includegraphics[width=\textwidth]{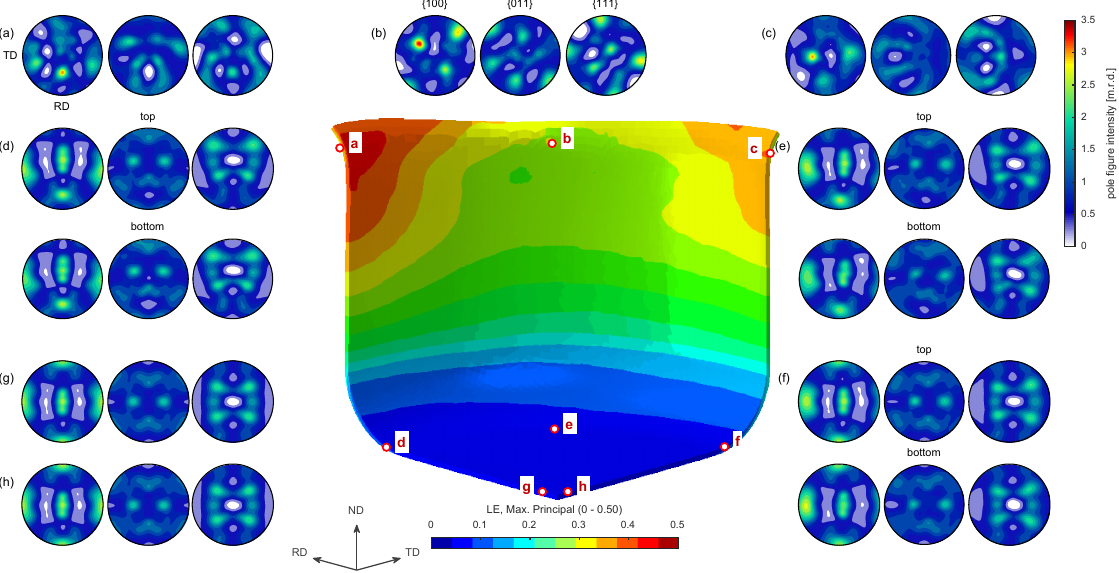}
  \caption{Local textures predicted by FE--ODMN with $N=8$ after springback. $\{100\}$, $\{011\}$, and $\{111\}$ pole figures at the rim (a--c), at the punch profile radius in the punch-side (top) and die-side (bottom) layers (d--f), and at the cup center (g, h), along the RD (a, d, g), 45$^\circ$ (b, e), and TD (c, f, h) rays. The cup is colored by the maximum principal logarithmic strain.}
  \label{fig:texture_cup}
\end{figure}

Figure~\ref{fig:texture_cup} shows the local textures predicted by the $N=8$ network after springback at points along the RD, 45$^\circ$, and TD rays. At the cup center, where the strain is small, the pole figures remain close to the initial texture. At the punch profile radius, the texture is also nearly unchanged, with only minor differences among the three rays and through the sheet thickness. At the rim, the texture has evolved considerably, because the material is compressed circumferentially in the flange and then bent and unbent over the die radius. The pole figures at the rim have higher intensities and different peak positions along the three rays, because the loading directions are oriented differently with respect to the initial rolling texture.

The texture-difference index of Eq.~\eqref{eq:Td}, taken with respect to the initial texture, quantifies this contrast. It reaches 0.28 to 0.42 at the rim and stays below 0.03 at the punch profile radius. The trend agrees with the self-consistent predictions of Barrett and Knezevic~\cite{barrett2019deep}. These local textures are part of the two-scale solution, since they are the material-node orientations stored at each integration point.

\subsection{Computational cost}
\label{sec:cost}

The cost has an offline part, the construction of the network, and an online part, the two-scale simulation. The offline part comprises the GSH texture compaction, the generation of the elastic training data, and the training. Each training sample requires six spectral solves of about 30~s each, and training takes 1, 4, and 22~min for $N=4$, 6, and 8 on a desktop computer with an Intel Core i9-13900 processor. This cost is incurred once for a given sheet material, whatever forming process is simulated afterward.

At the material-point level, the networks are about three to four orders of magnitude faster than the full-field spectral solver (Table~\ref{tab:rve_level}). A loading path that takes 16\,660~s with the spectral solver takes 2.8, 4.5, and 16.9~s with the $N=4$, 6, and 8 networks. At the component level, the complete cup-drawing simulations, including tool removal and springback, require 1\,440, 6\,290, and 25\,600 core-hours (Table~\ref{tab:cost}). The cost grows about fourfold with each fourfold increase in the number of material nodes.

For comparison, Barrett and Knezevic~\cite{barrett2019deep} reported about 4\,060 core-hours for an elasto-plastic self-consistent simulation of the same cup with 200 crystal orientations per integration point and 14\,564 elements. The present $N=6$ and 8 simulations require 1.5 and 6.3 times as many core-hours on a mesh with 1.9 times as many elements. Per element, these ratios become 0.8 and 3.3, although the two sets of simulations ran on different hardware.

\begin{table}[t]
  \centering
  \caption{Computational cost of the FE--ODMN cup-drawing simulations on the Forerunner~1 cluster (Intel Xeon Platinum 8480+ processors, Abaqus/Standard 2024). MPI stands for Message Passing Interface. A network solve is one evaluation of the ODMN at an integration point.}
  \label{tab:cost}
  \footnotesize
  \setlength{\tabcolsep}{3pt}
  \begin{tabular}{lccc}
    \toprule
    & $N=4$ & $N=6$ & $N=8$ \\
    \midrule
    Material nodes per integration point & 16 & 64 & 256 \\
    MPI processes (compute nodes) & 112 (1) & 112 (1) & 320 (4) \\
    Increments (drawing/unloading/springback)
      & 155/117/117 & 166/117/117 & 175/117/117 \\
    Network solves
      & $4.9\times10^{7}$ & $5.4\times10^{7}$ & $5.9\times10^{7}$ \\
    Wall time (h) & 12.9 & 56.1 & 80.0 \\
    Core-hours & 1\,440 & 6\,290 & 25\,600 \\
    \bottomrule
  \end{tabular}
\end{table}

\section{Discussion}
\label{sec:discussion}

The cup-drawing predictions depend on the friction coefficients and the contact stiffness, which were calibrated with the $N=6$ network (Section~\ref{sec:fe_model}) and kept unchanged for all three depths. Friction at the die and the holder mainly affects the drawing force, whereas friction at the punch affects thinning near the punch radius. The remaining thickness errors change little beyond $N=6$ and may come from the contact description, the crystal-plasticity law, or the RVE, and not from the network depth alone.

The crystal-plasticity model accounts for latent hardening but has no backstress and no mechanism for dislocation reversal. Heterogeneous stresses among the grains can produce a macroscopic Bauschinger effect, but the model may not reproduce the full reverse-loading behavior seen in the cyclic tests. This limitation could affect bending and unbending over the die radius and the subsequent springback. It can be addressed by extending the constitutive law at the material nodes, without any change to the network.

A deeper network represents the initial texture better and agrees better with the full-field simulations, but the discrepancies with the experiments do not vanish. For example, the $N=8$ network overpredicts the Lankford coefficient along RD with respect to the full-field solution, which in turn overpredicts the measurement. The remaining discrepancies stem from both the RVE and the network approximation. Barrett and Knezevic~\cite{barrett2019deep} also reported difficulties in predicting earing and showed that the ear heights are sensitive to the initial texture. The present model reproduces the TD ear height and the cup-height range but overpredicts the RD ears. Possible sources are errors in the directional plastic response, in the initial texture, and in the contact description. In addition, the quarter model forces opposite ears to have equal heights, and the uniform initial texture excludes texture variations across the blank. Both assumptions prevent the model from reproducing the asymmetry of the measured cup profile.

\section{Conclusions}
\label{sec:conclusions}

FE--ODMN has been applied to the cylindrical cup drawing of AA6022-T4 sheet, with an ODMN as a reduced-order RVE at each integration point, so that the macroscopic response is obtained from the microstructure without a macroscopic constitutive model. The main findings are as follows.

\begin{enumerate}
\item The sheet texture is the only microstructural input. The
  material-node orientations follow from GSH compaction, and the
  interaction directions are trained on linear-elastic data alone.
  No network parameter is fitted to the forming experiment. Only the
  contact parameters of the cup model are calibrated.

\item The $N=6$ and 8 networks reproduce the full-field RVE
  response within 3\,\% in flow stress and 5~MPa under load
  reversal, with speed-ups of about three to four orders of
  magnitude at the material point.

\item The $N=8$ network gives the peak punch force, the minimum
  wall thickness, and the cup-height range within 0.1\,\%, 3.4\,\%,
  and 0.7\,\% of the measurements, respectively. The ear positions
  agree with the experiment, although the RD ears remain too high.

\item The $N=6$ network already gives the punch force and the
  thinning, and $N=8$ improves the ear profile at four times the
  core-hours. A complete simulation takes 1\,400 to 26\,000
  core-hours, which makes two-scale crystal-plasticity analysis of
  a formed component feasible.

\item The local texture evolution follows from the material-node
  orientations at each integration point. The largest changes
  occur near the cup rim.
\end{enumerate}

FE--ODMN brings texture-dependent crystal plasticity and local texture evolution into component-scale forming simulations, which is a step toward integrated computational materials engineering of sheet-metal forming.

\section*{CRediT authorship contribution statement}

\textbf{Ting-Ju Wei}: Conceptualization, Methodology, Software, Validation, Data curation, Formal analysis, Investigation, Writing -- original draft, Visualization.
\textbf{Guo-Chi Li}: Conceptualization, Investigation, Writing -- review \& editing.
\textbf{Yi-Ting Lu}: Software, Investigation, Data curation, Formal analysis, Visualization, Writing -- review \& editing.
\textbf{Chuin-Shan Chen}: Conceptualization, Supervision, Funding acquisition, Formal analysis, Writing -- review \& editing, Project administration.

\section*{Declaration of competing interest}
The authors declare that they have no known competing financial interests
or personal relationships that could have appeared to influence the work
reported in this paper.

\section*{Acknowledgements}
This work is supported by the National Science and Technology Council, Taiwan, under Grant 114-2221-E-002-010-MY3. The authors gratefully acknowledge the computational resources and technical support provided by the National Center for Research on Earthquake Engineering (NCREE), National Institutes of Applied Research (NIAR), Taiwan, as well as by the NTUCE-NCREE Joint Artificial Intelligence Research Center and the National Center for High-performance Computing (NCHC). 

\section*{Data availability}
The datasets and trained models used in this study are available upon request from the corresponding author. 

\appendix
\setcounter{table}{0}\setcounter{figure}{0}
\renewcommand{\thetable}{A\arabic{table}}
\renewcommand{\thefigure}{A\arabic{figure}}
\renewcommand{\theequation}{A\arabic{equation}}\setcounter{equation}{0}
\section{Crystal-plasticity law of the material nodes}
\label{sec:cp}

The material nodes of the ODMN and the full-field reference simulations use the same phenomenological crystal-plasticity law of DAMASK~\cite{roters2019damask}, which is summarized here. For material node $i$, the deformation gradient is decomposed multiplicatively into an elastic part $\bm F_i^{\mathrm e}$ and a plastic part $\bm F_i^{\mathrm p}$, $\bm F_i=\bm F_i^{\mathrm e}\bm F_i^{\mathrm p}$. The initial crystal orientation $\mathbf g_i$ is represented by the passive rotation matrix $\bm G_i$, which sets the initial state
\begin{equation}
 \bm F_i^{\mathrm p}(0)=\bm G_i,\qquad
 \bm F_i^{\mathrm e}(0)=\bm G_i^{\mathsf T},\qquad
 \bm F_i(0)=\bm I,
 \label{eq:initial_orientation}
\end{equation}
where $\bm I$ is the identity tensor and the argument 0 denotes the initial time. The elastic response is Hooke's law with the cubic single-crystal stiffness $\C^{\mathrm c}$, which relates the second Piola--Kirchhoff stress $\bm S_i$ to the elastic Green--Lagrange strain $\bm E_i^{\mathrm e}$,
\begin{equation}
 \bm E_i^{\mathrm e}=\tfrac12\big[(\bm F_i^{\mathrm e})^{\mathsf T}
 \bm F_i^{\mathrm e}-\bm I\big],
 \qquad \bm S_i=\C^{\mathrm c}:\bm E_i^{\mathrm e},
 \qquad \bm P_i=\bm F_i^{\mathrm e}\bm S_i(\bm F_i^{\mathrm p})^{-\mathsf T}.
 \label{eq:hooke}
\end{equation}
Plastic flow takes place on the twelve $\{111\}\langle110\rangle$ slip systems $\alpha$ of the face-centered cubic lattice,
\begin{equation}
 \dot{\bm F}_i^{\mathrm p}(\bm F_i^{\mathrm p})^{-1}
 =\sum_{\alpha=1}^{12}\dot\gamma_i^\alpha\,\bm s^{\alpha}\otimes\bm n^{\alpha},
 \qquad
 \dot\gamma_i^\alpha=\dot\gamma_0
 \left|\frac{\tau_i^\alpha}{\xi_i^\alpha}\right|^{n}
 \operatorname{sgn}(\tau_i^\alpha),
 \label{eq:slip}
\end{equation}
where a superposed dot denotes the time derivative, $\dot\gamma_i^\alpha$ is the slip rate on system $\alpha$, $\bm s^{\alpha}$ and $\bm n^{\alpha}$ are the slip direction and the slip-plane normal, $\tau_i^\alpha=\bm S_i:(\bm s^{\alpha}\otimes\bm n^{\alpha})$ is the resolved shear stress, $\xi_i^\alpha$ is the slip resistance, $\dot\gamma_0$ is the reference slip rate, $n$ is the stress exponent, and $\operatorname{sgn}$ is the sign function. The slip resistances start from the initial value $\xi_0$ and evolve as
\begin{equation}
 \dot\xi_i^\alpha=h_0
 \left|1-\frac{\xi_i^\alpha}{\xi_\infty}\right|^{a}
 \operatorname{sgn}\!\Big(1-\frac{\xi_i^\alpha}{\xi_\infty}\Big)
 \sum_{\beta=1}^{12}h^{\alpha\beta}|\dot\gamma_i^\beta|,
 \label{eq:hardening}
\end{equation}
with the initial hardening modulus $h_0$, the hardening exponent $a$, the saturation resistance $\xi_\infty$, and the slip-system interaction matrix $h^{\alpha\beta}$~\cite{madec2017dislocation}. Each slip system hardens isotropically, without a backstress. The parameters in Table~\ref{tab:cp} are those of the DAMASK example configuration for AA6022-T4~\cite{damask_documentation}, which were fitted to the texture and tensile curves reported by Barrett and Knezevic~\cite{barrett2019deep}. The elastic constants are those of aluminum at room temperature from the same source. Only the reference slip rate $\dot\gamma_0$ differs between the two scales, because it is matched to the nominal strain rate of each simulation.
The current orientation of node $i$ is the rotation $\bm R_i$ in the polar decomposition $\bm F_i^{\mathrm e}=\bm R_i\bm U_i^{\mathrm e}$, where $\bm U_i^{\mathrm e}$ is the elastic right stretch tensor. The weighted orientations $\{(\bm R_i,w_i)\}$ define the local texture at each integration point.

\begin{table}[!ht]
  \centering
  \caption{Single-crystal constitutive parameters of AA6022-T4~\cite{damask_documentation}.
  The slip-interaction coefficients are listed in the order self,
  coplanar, collinear, Hirth, glissile (two types), and Lomer.}
  \label{tab:cp}
  \footnotesize
  \setlength{\tabcolsep}{4pt}
  \begin{tabular}{@{}lll@{}}
    \toprule
    Parameter & Symbol & Value \\
    \midrule
    Elastic constants
      & $C_{11}, C_{12}, C_{44}$
      & 106.9, 60.5, 28.4 GPa \\
    Initial slip resistance
      & $\xi_0$ & 76 MPa \\
    Saturation slip resistance
      & $\xi_\infty$ & 266 MPa \\
    Initial hardening modulus
      & $h_0$ & 1\,020 MPa \\
    Hardening exponent
      & $a$ & 3.7 \\
    Stress exponent
      & $n$ & 20 \\
    Reference slip rate, RVE level
      & $\dot{\gamma}_0$ & $10^{-3}$ s$^{-1}$ \\
    Reference slip rate, cup model
      & $\dot{\gamma}_0$ & 1 s$^{-1}$ \\
    
Slip-interaction coefficients
  & $h^{\alpha\beta}$
  & \begin{tabular}[c]{@{}c@{}}
      $[1,\,1,\,5.123,\,0.574,$ \\
      $1.123,\,1.123,\,1]$
    \end{tabular} \\

    \bottomrule
  \end{tabular}
\end{table}

\bibliographystyle{elsarticle-num}
\bibliography{references}

\end{document}